\documentclass[preprint,12pt]{elsarticle}

\usepackage{amssymb}
\usepackage{amsmath}
\usepackage{graphicx}
\usepackage{xcolor}
\newcommand{\JS}[1]{{\textcolor{black}{#1}}}

\journal{Tribology International}

\begin{document}

\begin{frontmatter}

\title{The slope of the friction law of hertzian-asperity--based metainterfaces has a finite positive lower bound}

\author{Julien Scheibert} %% Author name

\affiliation{organization={CNRS, Ecole Centrale de Lyon, ENTPE, LTDS, UMR5513, 69134 Ecully}, country={France}}

\begin{abstract}
Metainterfaces can realize specified evolutions of their friction force as a function of the confining normal force (friction law), thanks to the design of the individual radii and heights of a population of independent hertzian asperities. However, not all friction laws are achievable. Here I show that, contrary to a suggestion from the literature, the slope of the friction law has a finite positive lower bound. \JS{This result is useful to identify friction laws that are not accessible to metainterfaces.}
\end{abstract}

\end{frontmatter}

Since Greenwood and Williamson~\cite{greenwood_contact_1966}, a whole branch of rough contact mechanics has been using models based on independent hertzian asperities to evaluate the real contact area, $A_0$, and its evolution as a function of the normal force, $P$, applied on the interface. Although spectral models are better-suited for the multiscale roughness of natural or engineering surfaces~\cite{vakis_modeling_2018}, the emerging concept of frictional metainterfaces~\cite{aymard_designing_2024,zeka_normal_2026,fu_automated_2026} makes a deliberate use of asperity models to simplify the design of architectured interfaces having specified contact area and friction force, $F$.

The friction model of~\cite{aymard_designing_2024} \JS{(see Eqs.~\ref{Eq:F}-\ref{Eq:P} below)} has recently been used in~\cite{mouton_friction_2026} to create a database for the relationship between hertzian-asperity--based topographies and friction laws, $F(P)$. The authors of~\cite{mouton_friction_2026} illustrate the contents of the database with select friction laws having various qualitative shapes. One of them has a saturating part, where the friction remains constant after an initial increasing part. Such a vanishing slope is impossible within the friction model underlying the metainterfaces of~\cite{aymard_designing_2024}, which I demonstrate below.

\section{Lower limit of the slope of the friction law}

\begin{figure}[h!]
\centering
\includegraphics[width=0.8\textwidth]{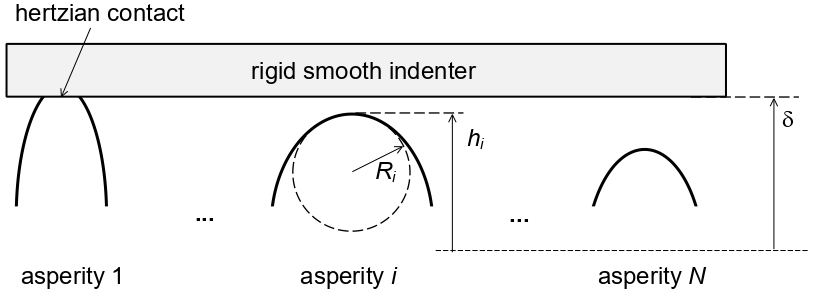}
\caption{Sketch of the metainterface model. $N$ linear elastic parabolic asperities, each with a different curvature radius ($R_i$) and height ($h_i$) of its apex, create independent hertzian contacts with a rigid smooth indenter of altitude $\delta$.}
\label{Fig:Asperities}
\end{figure}

The friction model of~\cite{aymard_designing_2024} is based on $N$ parabolic linear elastic asperities, each with a different radius of curvature, $R_i$, and height, $h_i$, \JS{of its apex (Fig.~\ref{Fig:Asperities})}. Each microcontact is supposed to obey Hertz's behaviour~\cite{barber_contact_2018}. They are assumed independent, so that the total forces are just sums over the $N$ asperities:
\begin{align}
F(\delta) &= B \sigma A_0(\delta) =  \pi B \sigma \sum_{i=1}^{N} R_i (h_i - \delta) H(h_i - \delta),\label{Eq:F}\\
P(\delta) &= \frac{4}{3} E^* \sum_{i=1}^{N} \sqrt{R_i} (h_i - \delta)^{3/2} H(h_i - \delta),\label{Eq:P}
\end{align}
where $\delta$ is the altitude of the rigid and smooth indenting plane, $B \sigma$ is the proportionality factor that relates friction to the contact area ($\sigma$ is the friction stress and $B$ quantifies shear-induced area reduction~\cite{sahli_evolution_2018}), and $H$ is the Heaviside function ($H(x)$=1 for $x\ge0$ and 0 otherwise). 

Differentiating both $F$ and $P$ with respect to $\delta$ and dividing one by the other, we can express the slope of the friction law, $F(P)$, as:
\begin{equation}
\frac{dF}{dP}(\delta) = \frac{\pi}{2}\frac{B \sigma}{E^*} \frac{\sum_{i=1}^{N} R_i H(h_i - \delta)}{\sum_{i=1}^{N} \sqrt{R_i} \sqrt{h_i - \delta} H(h_i - \delta)}.\label{slope}
\end{equation}
In practice, a given metainterface is characterized by an interval of radii, $[R_{min};R_{max}]$, and an interval of heights, $[h_{min};h_{max}]$. With those values, we can estimate a lower bound for the slope $\frac{dF}{dP}$ as follows. A lower bound of the sum in the numerator in Eq.~\ref{slope} is $R_{min} \sum_{i=1}^{N} H(h_i - \delta)$, while an upper bound of the sum in the denominator is $\sqrt{R_{max}}\sqrt{h_{max}-\delta} \sum_{i=1}^{N} H(h_i - \delta)$. A lower bound of the slope of the friction law can thus be defined as:
\begin{equation}
\frac{dF}{dP}(\delta) \ge  \frac{\pi}{2}\frac{B \sigma}{E^*} \frac{R_{min}}{\sqrt{R_{max} (h_{max}-\delta)}}.\label{Eq:LB}
\end{equation}
This expression, \JS{valid for any asperity height distribution}, shows that the lower bound, although being a decreasing function of the indentation, $h_{max}-\delta$, vanishes only asymptotically. Thus, in all practical cases, the slope has a strictly positive finite value. As a consequence, the friction law of hertzian-asperity--based metainterfaces is a monotonous, strictly increasing function and cannot exhibit sustained saturation within a range of finite indentation.

\section{Application to Fig.~A.11d of~\cite{mouton_friction_2026}}\label{sec:saturation}

\begin{figure}[ht]
\centering
\includegraphics[width=0.8\textwidth]{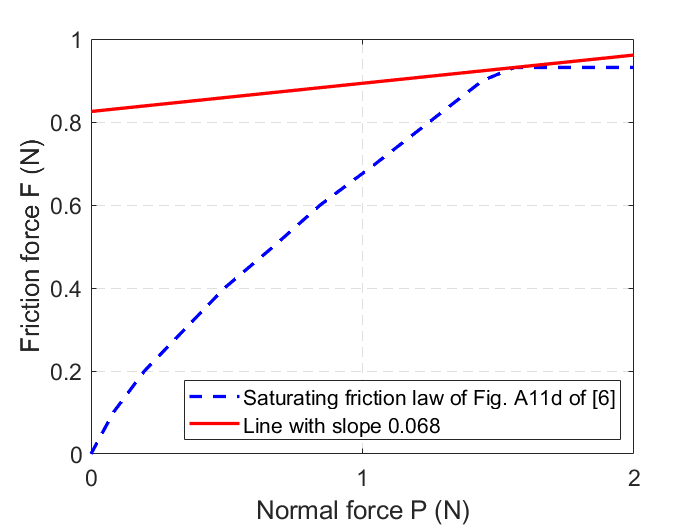}
\caption{Dashed blue curve: redrawing of the saturating friction law shown in Fig.~A.11d of~\cite{mouton_friction_2026}. Red line: straight line with slope 0.068 (conservative value of the lower limit of the slope when using parameters taken from~\cite{mouton_friction_2026}, see text for details).}
\label{Fig:MoutonvsSlope}
\end{figure}

\JS{Figure~\ref{Fig:MoutonvsSlope} reproduces in dashed blue line the saturating friction law shown in Fig.~A.11d of~\cite{mouton_friction_2026}. Unfortunately, }the indentation $\delta$ at which saturation starts is not explicitly provided, which makes it impossible to apply Eq.~\ref{Eq:LB} directly. So we estimate here a very conservative value of the slope's lower bound by replacing, in Eq.~\ref{Eq:LB}, $\delta$ with $\delta_{min}$, the smallest considered altitude of the indenting plane. Using all parameter values from~\cite{mouton_friction_2026} ($B$=0.85, $\sigma$=0.4\,MPa, $E^*$=1.36\,MPa, $h_{max}$=300\,$\mu$m, $\delta_{min}$=0\,$\mu$m, and $R_{min}$=60\,$\mu$m and $R_{max}$=400\,$\mu$m estimated from the radius distribution of Fig.~A.11d), and considering that the friction model of Eqs.~\ref{Eq:F}-\ref{Eq:P} is used, the slope cannot be smaller than about 0.068. \JS{As illustrated in red line in Fig.~\ref{Fig:MoutonvsSlope},} this (underestimated) lower bound of the slope would correspond to an increase of $F$ of about 0.03\,N over the range of the saturating part of the friction law ($P$ in the approximate range [1.56 ; 2.00]\,N), incompatible with the curve plotted in Fig.~A.11d of ~\cite{mouton_friction_2026}.

\section{Discussion}

\JS{So far, the experimentally demonstrated metainterfaces~\cite{aymard_designing_2024,fu_automated_2026} have mostly used asperities with a single radius of curvature, so that $R_\text{min}=R_\text{max}=R$. They have also been indented down to $\delta=0$. In these conditions, the smallest slope predicted by Eq.~\ref{Eq:LB} is $\frac{\pi}{2}\frac{B \sigma}{E^*} \sqrt{\frac{R}{h_\text{max}}}$, irrespective of the asperity height distribution. Consistently, it is smaller than the minimum asymptotic slopes (when $\delta \rightarrow 0$) found for quasi-linear friction laws (height distribution following a truncated exponential) in~\cite{aymard_designing_2024}: $\frac{3\pi}{4}\frac{B \sigma}{E^*} \sqrt{\frac{R}{h_\text{max}}}$. It is a fortiori also smaller than the minimum asymptotic slope of proportional friction laws (height distribution being a triangular-tailed exponential) identified in~\cite{fu_automated_2026}: $1.36\times\frac{3\pi}{4}\frac{B \sigma}{E^*} \sqrt{\frac{R}{h_\text{max}}}$.}

\JS{In~\cite{fu_automated_2026}, bilinear friction laws were introduced, where $F(P)$ has a first linear segment at low normal forces, and a second linear segment at higher normal forces, with the slope of the second segment being  potentially smaller than in the first. Such a smaller slope was achieved using asperities with two possible radii of curvature, $R$ and $3R$, which can also be interpreted with Eq.~\ref{Eq:LB}: by using two different radii, $\frac{R_\text{min}}{\sqrt{R_\text{max}}}$ is reduced compared to when a single radius is used, and thus the lower limit of the slope is also reduced. This property presumably explains why achieving smaller slopes in the second segment is easier with two radii than with only one.}

\begin{figure}[h!]
\centering
\includegraphics[width=0.8\textwidth]{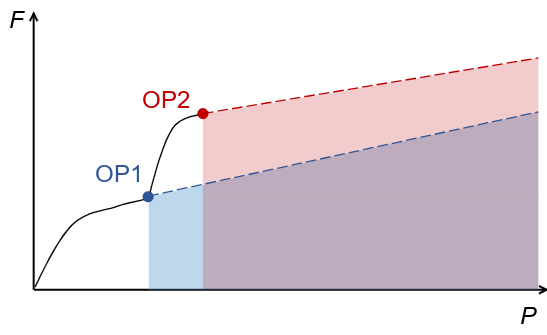}
\caption{Sketch of a friction law (black curve) defined by successive operating points (OP). Dashed colored lines: lines starting from each OP (blue for OP1, red for OP2) with a slope equal to the lower limit at the corresponding $\delta$. Blue (red) shaded area: forbidden area when the friction law has reached the blue (red) OP.}
\label{Fig:forbidden}
\end{figure}

\JS{More generally, Eq.~\ref{Eq:LB} has the potential to facilitate the design of metainterfaces. It provides a sufficient condition to decide whether a target friction law is unphysical: if the slope of the target law is at any point smaller than its lower limit, then no design solution can be found. Beyond showing that saturating laws are impossible (see section~\ref{sec:saturation}), it can also be used for instance when specifying a target friction law based on successive operating points $(P_i,F_i)$ (like those presented in~\cite{aymard_designing_2024}). The lower limit can be applied at each operating point to define a forbidden region in the $\{P,F\}$ plane, which cannot be accessed starting from the current operating point and where no further OP can be placed. This is illustrated in Fig.~\ref{Fig:forbidden}.}

\section{Conclusion}
\JS{An analytical, strictly positive lower bound of the slope of the friction law of hertzian-asperity--based metainterfaces has been derived. It constitutes a conservative but simple sufficient condition to assess whether a given target friction law cannot be obtained through designing a metainterface.}

\section*{Acknowledgements}
The  author  is indebted  to  the  Carnot  institute  Ingénierie@Lyon,  labelled  by  the  French National Research Agency (ANR), for its support and funding. He thanks Davy Dalmas and Li Fu for discussions.

\bibliographystyle{elsarticle-num} 
\bibliography{CommentMoutonRevised}

\end{document}